\documentclass[AMA,STIX2COL]{MRM}
\articletype{Research Article}%
\usepackage{amssymb}
\usepackage{float}
\usepackage{setspace}
\usepackage{xr}
\usepackage{tocloft}
\usepackage{marginnote}
\usepackage{xcolor}

\newcommand{\putmarginpar}[2]{#2}

\received{26 April 2016}
\revised{6 June 2016}
\accepted{6 June 2016}
\newcommand{\bs}{\boldsymbol}
\newcommand{\mc}{\mathcal}

\newcommand{\mbb}{\mathbb}

\newcounter{suppfig}
\renewcommand{\thesuppfig}{S\arabic{suppfig}} 
\newcommand{\suppfigcaption}[2]{
  \refstepcounter{suppfig} 
  \par\noindent\textbf{Figure \thesuppfig.} #2 
  \label{suppfig:#1} 
  \addcontentsline{exp}{listsuppfig}{\protect\numberline{\thesuppfig} #2}
}
\newcounter{supptab}
\renewcommand{\thesupptab}{S\arabic{supptab}} 
\newcommand{\supptabcaption}[2]{
  \refstepcounter{supptab} 
  \par\noindent\textbf{Table \thesupptab.} #2 
  \label{supptab:#1} 
  \addcontentsline{exp}{listsupptab}{\protect\numberline{\thesupptab} #2}
}

\usepackage{multicol,zref-savepos,color,xparse}

\begin{document}

\title{Unified Motion Correction and Reconstruction for Accelerated Zero Echo Time MRI}


\author[1]{Yan Chen}{\orcid{0009-0003-6886-1886}}

\author[2]{Curtis A. Corum}{\orcid{0000-0002-4994-6206}}

\author[3,4]{James H. Holmes}{\orcid{0000-0002-0936-3700}}

\author[1]{Mathews Jacob}{\orcid{0000-0001-6196-3933}}

\authormark{Chen \textsc{et al}}

\address[1]{\orgdiv{Electrical and Computer Engineering}, \orgname{University of Virginia}, \orgaddress{\state{Virginia}, \country{U.S.}}}

\address[2]{\orgdiv{Champaign Imaging LLC}, \orgaddress{\city{St. Paul}, \state{Minnesota}, \country{U.S.}}}

\address[3]{\orgdiv{Department of Radiology}, \orgname{University of Iowa}, \orgaddress{\state{Iowa}, \country{U.S.}}}

\address[4]{\orgdiv{Biomedical Engineering and Hawk-IDDRC}, \orgname{University of Iowa}, \orgaddress{\state{Iowa}, \country{U.S.}}}

\corres{Mathews Jacob\\ \email{wzr8sn@virginia.edu}}


\finfo{The work is supported by R01-AG067078 and R01-EB019961. The data acquisition was supported by R43MH122028 and was acquired on an MRI instrument funded by S10OD025025. The University of Iowa receives research support from GE Healthcare. University of Virginia receives support from Siemens Healthineers and Canon Medical Systems.}

\abstract{
\vspace{-1em}
\section{Purpose}
To develop an accelerated Zero Echo Time (ZTE) MRI reconstruction framework that jointly estimates the 3D image and rigid-motion parameters, while compensating for missing k-space center data and angular undersampling artifacts within a regularized maximum a posteriori (MAP) formulation.
\section{Methods}
We formulate the joint recovery as a MAP estimation, termed Energy-based Motion Correction (EnMoCo), which is solved by alternating between k-space rigid-motion updates (trajectory rotation and phase modulation) and conjugate-gradient data-consistency updates. Regularization is provided by a pre-trained Multi-Scale Energy-based model (MuSE), implemented in a data-efficient 2.5D form by combining slice-wise energies from two orthogonal views to reduce 3D training demands.  
\section{Results}
We demonstrate the effectiveness of the proposed method on in vivo 3D ZTE data and simulated data to replicate more challenging hardware scenarios, including dead time gaps spanning 1.5--5 samples, rotations up to $\pm17^\circ$, translations up to $\pm10$~mm, and a range of scan times (153 to 611 seconds). In simulations, the proposed method improves both PSNR and SSIM relative to CG-SENSE and Autofocusing (e.g., $\sim$10--15 dB higher PSNR for gap $\geq$3.5), and reduces rotational motion error in accelerated experiments. Experimental reconstructions show reduced motion blur and inconsistencies in the tissue contrast compared with the same baselines across larger dead time gaps of 3--4.5 samples.

\section{Conclusion}
A unified MAP formulation with k-space motion modeling and an explicit energy-based prior can improve reconstruction quality for motion-corrupted, accelerated ZTE relative to standard reconstruction and motion-correction baselines evaluated in this work. 
}

\wordcount{5000}
\maketitle 


\section{Introduction}\label{sec1}
\vspace{-1em}

Zero echo-time (ZTE) MRI is actively explored\cite{aydingoz2022zero,ljungberg2021silent,weiger2011mri,ljungbergMotionCorrectedSilent2022a, wiesinger2022zero,ljungberg2020silent} for its near-silent and time-efficient 3D isotropic image acquisition. Silent ZTE relies on radial acquisitions, where gradient directions are incremented between repetitions instead of being ramped up and down repeatedly \cite{wiesinger2016zero}. Slow gradient increments reduce scan noise compared to regular radial acquisitions, enhancing patient comfort, especially for pediatric subjects and individuals with autism spectrum disorder (ASD), Down syndrome, or dementia. Despite its advantages, ZTE introduces several intrinsic reconstruction challenges. Because the readout gradient is active during RF excitation and no prewinder lobes are used, the acquisition inherently misses k-space samples near the center~\cite{wiesinger2022zero}.  Although reducing the readout bandwidth can decrease the dead time gap, it may also introduce chemical-shift artifacts and soft-tissue intensity nonuniformity. These trade-offs motivate high-bandwidth ZTE protocols that rely on algorithmic strategies to recover missing data\cite{engstrom2020phase}. Although silent ZTE improves comfort and reduces the likelihood of motion during scanning, the prolonged acquisition can make it challenging for subjects to remain still throughout the scan, particularly for individuals with ASD. Hence, the simultaneous presence of missing central k-space samples and subject motion further complicates accurate ZTE MRI reconstruction.

Traditionally, missing k-space samples are filled using additional acquisitions designed to sample the central k-space region (e.g., WASPI\cite{wu2007waspi}, PETRA\cite{grodzki2012ultrashort}, and HYFI\cite{froidevaux2021hyfi}). A key limitation of such hybrid strategies is that data from the center of k-space may differ from data at the outer regions collected using the ZTE acquisition, due to either motion or magnetization preparation state. To avoid separate central acquisitions, several purely computational infilling methods have been proposed\cite{froidevaux2018fillinga,corum2020robust,stoch2005missing, huoLowrankIterativeInfilling2025,wood2021zte}. These approaches exploit finite image support\cite{kuethe1999transforming,corum2020robust,stoch2005missing} and/or parallel imaging\cite{wood2021implicit} to estimate the missing samples directly from the acquired data. However, these approaches are proposed for motion-free settings. Applying these methods to motion-corrupted ZTE data poses additional challenges due to inconsistency across spokes. Moreover, some infilling strategies, such as Low Rank\cite{huoLowrankIterativeInfilling2025}, are computationally expensive when the dead time gap is large, making them difficult to integrate with iterative motion estimation and correction.  

 A common approach for motion estimation is to recover low-resolution navigator volumes from the central k-space signal\cite{welch2004self,kecskemeti_robust_2018,wallace2024rapid,vaillant2013retrospective}, followed by the estimation of rigid motion parameters from the navigator volumes\cite{kecskemeti_robust_2018} by image registration in the image domain. In addition to being computationally expensive, approaches using image domain navigators have a trade-off: smaller bins improve the temporal resolution of motion estimates, while larger bins are often needed to produce high-SNR navigators for accurate registration and reliable motion parameters. An alternative approach is Autofocusing (AF) \cite{atkinson1999automatic,loktyushin2013blind,kuzmina2022autofocusing+}, where rigid motion parameters are iteratively adjusted to optimize an image sharpness metric (e.g., entropy) or data consistency that exploits parallel imaging redundancy\cite{haskell2018targeted,cordero-grande2020motioncorrected,al-hajhemidiIMMoCoSelfsupervised2024,wu2025moner,chen2023dfm,liu2026samoco}. These approaches are not directly applicable for accelerated ZTE, as dead time gaps and acceleration enlarge the null-space of the encoding operator, and because motion and undersampling errors are strongly coupled, making motion-compensated recovery ill-posed. For ZTE motion correction, MERLIN\cite{ljungbergMotionCorrectedSilent2022a} adopts registration between navigator images using coil sensitivity maps provided by WASPI, which resolves the missing samples issue using an additional low-resolution scan and is constrained by the SNR of navigator images.

Regularized maximum a posteriori (MAP) reconstruction provides a natural framework for ZTE, where multiple degradations can be handled jointly, while a strong image prior can make the problem well-posed. While learning-based priors have recently improved MRI reconstruction, many of the widely used formulations are difficult to deploy in the 3D ZTE setting. For example, end-to-end unrolled architectures \cite{varnet,aggarwal2018modl,monga2021algorithm} can be memory-demanding for 3D radial applications, because reconstruction cannot be decomposed into independent 2D slices. Plug-and-play reconstruction with pre-trained denoisers \cite{venkatakrishnan2013plug,ahmad2020plug} alleviates some of these practical constraints, but convergence guarantees often require conservative assumptions on the denoiser (e.g., Lipschitz continuity or nonexpansiveness), which may reduce reconstruction quality \cite{hurault2021gradient,chand2024multi}. Although self-supervised approaches do not require training data, their reconstruction process is often time-consuming\cite{van2026time,chen2025accelerating}. Custom-trained deep learning models for motion compensation \cite{haskell2019network,johnson2019conditional} have also been developed using paired motion-free and motion-corrupted images or k-space data\cite{singh2024data,al-hajhemidiIMMoCoSelfsupervised2024,dabrowski2024sismik}. Recently, diffusion-based approaches have also been proposed for 2D image recovery and motion correction\cite{levac2024accelerated,chen2025joint}, using diffusion posterior sampling (DPS) together with motion compensation applied to intermediate noisy latents. Translating these benefits to ZTE is challenging, primarily due to limited training data and the substantial computational burden of diffusion inference in 3D.

 We develop a regularized reconstruction framework for 3D ZTE acquisitions affected by (i) a missing k-space center, (ii) subject motion, and (iii) undersampling from accelerated scans. We cast image recovery as a regularized MAP estimation problem, solving for the image jointly with nuisance variables, including rigid-motion parameters and unmeasured k-space samples (both the central gap and undersampled locations), within a unified optimization scheme. For regularization, we use the Multi-Scale Energy-based model (MuSE)\cite{chand2024multi}, which provides an explicit energy (negative log-prior) parameterized by a compact CNN trained via multi-scale denoising score matching. The MuSE prior used in this work is algorithmically unrelated to MUSE (Multiplexed Sensitivity-Encoding\cite{chen2013robust}) and AMUSE\cite{guhaniyogi2016motion}, which are SENSE-based reconstruction methods for multishot diffusion MRI. The explicit prior distinguishes MuSE from plug-and-play approaches based on implicit denoisers \cite{venkatakrishnan2013plug,ahmad2020plug,haskell2019network} and allows the use of optimization algorithms with established convergence properties under standard assumptions\cite{hurault2021gradient}. We compute a MAP estimate using a fast majorization-minimization strategy as in \citenum{chand2024multi}. From a practical standpoint, MuSE-based MAP inference is typically less computationally demanding than diffusion posterior sampling methods, which can make it more tractable for 3D settings. In addition, the compact CNN prior can be trained with relatively limited data, which is helpful in our constrained-data regime. To further reduce training requirements, we adopt a 2.5D construction in which the 3D prior is approximated as a sum of slice-wise energies computed from two orthogonal views. For motion, we model rigid-body effects directly in k-space through trajectory rotations and the corresponding phase modulations, enabling motion compensation within the data-consistency step rather than relying on reconstructions from binned navigator images. This approach is intended to reduce dependence on binning choices and the associated trade-off between temporal resolution and navigator fidelity. The resulting algorithm alternates between motion-parameter estimation, data-consistency updates, and MuSE-based regularization. \putmarginpar{R1.C1.}{We term the proposed method EnMoCo (\textbf{En}ergy-based \textbf{Mo}tion \textbf{Co}rrection).}

\section{Methods}
We propose a regularized MAP reconstruction framework for 3D ZTE that jointly estimates the image, rigid-motion parameters, and unmeasured k-space samples. The method uses an explicit MuSE energy prior \cite{chand2024multi} in a data-efficient 2.5D form, alternating majorization-minimization image updates with k-space motion compensation.

\subsection{Forward Model with Rigid Motion}
We denote the multichannel MRI measurements of an image $\bs x$ at a time $\tau$ and the k-space location $\mathbf k_{\tau}$ as 
\begin{equation}
    \bs y_{\tau} = \mc F_{\mathbf k_{\tau}}~ \bs x+\bs n_{\tau},
\end{equation}
where $\mc F_{\mathbf k}$ denotes the three-dimensional Fourier transform with coil sensitivity weighting. Here, $\bs n_{\tau}$ and $\bs y_{\tau}$ represent the noise and measurements in k-space, respectively.

We assume the brain to be a rigid body and model the motion at time $\tau$ by $\bs \theta_{\tau}$ and $\bs t_{\tau}$, which are the three rotation and three translation parameters. The measured k-space data at a specific time point $\tau$ thus correspond to the Fourier transform of the rotated and translated version $\bs x_{\tau}'$, where 
\begin{equation}
    \bs x_{\tau}^{\prime} = \mathbf R_{\bs \theta_{\tau}}\bs x-\bs t_{\tau}.
\end{equation}

We concatenate the rotation parameters $\bs \theta \in \mathbb{R}^{T \times 3}$ and translation parameters $\bs t \in \mathbb{R}^{T \times 3}$ along the feature dimension to obtain the joint motion parameter matrix $\bs \xi \in \mathbb{R}^{T \times 6}$ across all $T$ time points, as shown below:
\begin{equation}
    \bs \xi = \begin{bmatrix} \bs \theta & \bs t \end{bmatrix} \in \mathbb{R}^{T \times 6}.
\end{equation}

Using the properties of the Fourier transform, the operation $\mc F_{\mathbf k_{\tau}} \bs x_{\tau}^{\prime}$ can be expressed as:

\begin{equation}
    \mc F_{\mathbf k_{\tau}} \,\bs x_{\tau}^{\prime} = {\bs{\mc A}_{\bs\xi_{\tau}}}\bs{x}.\
\end{equation}
This reformulation allows us to model the k-space signal of the rotated image $\bs{{x}}^{\prime}_{\tau}$ as the modified forward model $\bs{\mc A}_{\bs\xi_{\tau}}$ applied to the static image $\bs{x}$. In particular, a rotation $\mathbf R_{\theta}$ in the image domain corresponds to the rotation by the same angle in k-space. Similarly, a phase modulation in k-space accounts for a translation $\bs t$ in the image domain. The above properties allow us to account for motion exclusively in k-space, without having to generate intermediate images. 

\subsection{Joint Recovery of Motion \& Image }

We propose to jointly estimate the image $\bs{{x}}$ and the 3D motion parameters $\bs{\xi}$ by minimizing the cost function $\mc{L}(\bs {x};\bs \xi)$:
\begin{equation}
\label{eq:obj}
\mc{L}(\bs {x};\bs \xi) \;=\; \tfrac{1}{2\sigma^2}\,\|\bs y- \bs{\mc A}_{\bs \xi}\bs {x}\|_2^2 \;+\;\mc{R}_{\bs w}(\bs {x}),
\end{equation}
where $\mathcal{R}_{\boldsymbol{w}}:\mathbb{C}^{M_x\times M_y\times M_z}\to \mathbb{R}$ is an image domain regularizer, represented by the 3D MuSE \cite{chand2024multi}, and $\sigma^2$ is the variance of the measurement noise. To train with limited data, we adopt a 2.5D strategy, where the 3D prior is the sum of the MuSE energies of the 2D slices. The motion parameters $\bs \xi$ are estimated simultaneously during the reconstruction of the image volume $\bs x$. 

\paragraph*{2D MuSE regularizer}
Let $\bs\nu \in \mathbb{C}^{M_x \times M_y}$ denote a 2D slice in the 3D volume. MuSE models the prior distribution of $\bs\nu$ as
\begin{equation}
\label{prior}
p_{\bs{w}}(\bs\nu) \;=\; \frac{1}{Z_{\bs{w}}} \exp\!\big(-r_{\bs{w}}(\bs\nu)\big),
\end{equation}
where $Z_{\bs{w}}$ is the partition function (normalization constant), $r_{\bs{w}}(\bs{\nu})$ denotes the energy of a 2D image $\bs{\nu}$, and $\bs{w}$ denotes the parameters of the energy model $r_{\bs{w}}(\cdot): \mathbb{C}^{M_x \times M_y} \to \mathbb{R}$. The energy is parameterized as\cite{chand2024multi}
\begin{equation}
\label{eq:muse_energy}
r_{\bs{w}}(\bs\nu) \;=\; \frac{1}{2\varsigma_0^2}\,\big\|\bs\nu - \mc{D}_{\bs{w}}(\bs\nu)\big\|_2^2,
\end{equation}
where $\mc{D}_{\bs{w}}(\cdot): \mathbb{C}^{M_x\times M_y}\rightarrow \mathbb{C}^{M_x\times M_y}$ is a compact DRUNet-style CNN, and $\varsigma_0>0$ is a fixed small constant.

MuSE learns $\bs{w}$ using multi-scale denoising score matching. In particular, for noise levels $\varsigma$ drawn from a prescribed schedule and Gaussian perturbations $\bs z \sim \mc{N}(\bs 0,\bs I)$, the parameters are obtained by minimizing

\begin{equation}
\label{eq:score_matching}
        \bs{w}^* = \arg \min_{\bs{w}} \mbb E_{\varsigma} \big( \mbb E_{\bs \nu } \mbb E_{\bs z}\big [\|\nabla_{\bs \nu } \mathbf {r}_{\bs{w}} (\bs \nu  + \varsigma \bs z) - \varsigma \bs z \|_2^2\big ]  \big),
    \end{equation}

\noindent which encourages $\nabla_{\bs\nu}r_{\bs{w}}$ to match the Gaussian corruption process across multiple noise scales. Training across multiple $\varsigma$ values has been reported to improve robustness during optimization \cite{chand2024multi}. Consequently, the learned CNN $\mathcal{D}_{\mathbf{w}}(\cdot)$ robustly removes noise across multiple acceleration rates and varying degrees of blur.

\paragraph{2.5D regularization using the 2D model}
Let $\bs{x} \in \mathbb{C}^{M_x\times M_y\times M_z}$ denote the 3D volume. We first define an energy by summing MuSE slice energies along the $z$ direction:
\begin{equation}
\mc{R}^z_{\bs{w}}(\bs{x}) \;=\; \sum_{s_z=1}^{M_z}\mathbf {r}_{\bs{w}}\!\big(P_{s_z}\bs{x}\big),
\end{equation}
where $P_{s_z}$ is a linear slice selection operator to extract the $x$--$y$ slice at index $z=s_z$. $\mc{R}_{\bs{w}}^z$ denotes the energy of a 3D volume as a summation of 2D energies along $z$ direction, $\mathcal{R}^z_{\boldsymbol{w}}(\cdot): \mathbb{C}^{M_x \times M_y \times M_z} \to \mathbb{R}$. This construction allows training to leverage all 2D slices extracted from each 3D volume, which can reduce the demand for fully independent 3D training examples. 

To mitigate view-dependent artifacts (i.e., artifacts in planes orthogonal to the training orientation), we form a 2.5D prior by averaging energies computed from two complementary views and obtain $\mathcal{R}_{\boldsymbol{w}}(\cdot): \mathbb{C}^{M_x \times M_y \times M_z} \to \mathbb{R}$:
\begin{equation}
\label{3dmuse}
\mc{R}_{\bs{w}}(\bs{x}) \;=\; \frac{1}{2}\Big(\mc{R}^z_{\bs{w}}(\bs{x}) + \mc{R}^y_{\bs{w}}(\bs{x})\Big),
\end{equation}
where $\mc{R}^y_{\bs{w}}(\bs{x})=\sum_{s_y=1}^{M_y}r_{\bs{w}}({P}_{s_y}\bs{x})$ sums energies of $x$--$z$ slices extracted at $y=s_y$.

\subsection{Maximum-a-posteriori (MAP) optimization}
Given the prior $\mc{R}_{\bs{w}}(\bs{x})$, we utilize the Majorization-minimization (MM) optimization scheme to solve the cost function in Eq.~\eqref{eq:obj}. 
\begin{equation}
\label{eq:mm_rho}
\begin{aligned}
    \{\bs {x}^{n+1},\bs \xi^{n+1}\} &= \arg \min_{\bs {x},\bs\xi} \frac{1}{2\sigma^2}\|\bs y-\bs{\mc A}_{\bs \xi}\bs {x}\|_2^2  +\mc{R}_{\bs{w}}(\bs {x}^n) \\
    &+ \nabla_{\bs {x}}\mc{R}_{\bs{w}} (\bs {x}^n)(\bs {x}-\bs {x}^n) + \frac{L}{2}\|\bs {x}-\bs {x}^n\|^2_2 ,\\
\end{aligned}
\end{equation}
where $L$ is the Lipschitz constant of the score network, and $n$ is the index of the iteration. MM reformulates the optimization into an alternating minimization scheme, which alternates between motion update using steepest descent:
\begin{eqnarray}\label{motionupdate}
    \bs \xi^{n+1} &=& \bs \xi^n - \alpha \nabla_{\bs \xi}\tfrac{1}{2\sigma^2}\,\|\bs y-\bs{\mc A}_{\bs \xi^n}\bs {x}^{n}\|_2^2 ,
    \label{eq:motionupdate}
\end{eqnarray}
and image update using the Conjugate Gradient (CG) algorithm:
\begin{eqnarray}\nonumber
       \bs {x}^{n+1} &=&\left(\frac{\bs{\mc A}_{\bs \xi^{n+1}}^H \bs{\mc A}_{\bs \xi^{n+1}}}{\sigma^2}+L\bs I\right)^{-1}\left(\frac{1}{\sigma^2}\bs{\mc A}_{\bs \xi^{n+1}}^H\bs y+L \hat{\mathbf{x}}^{n}\right),\\\label{eq:cgupdate}
\end{eqnarray}
where $\hat{\mathbf{x}}^{n}$ is the denoised version of $\bs x^n$:
\begin{equation}
    \hat{\mathbf{x}}^{n} = \left(\bs {x}^n-\frac{1}{L}\nabla_{\bs {x}}\mc{R}_{\bs{w}} (\bs {x}^n)\right),
    \label{eq:denoise}
\end{equation}
 obtained by subtracting the MuSE score. The algorithm implicitly recovers missing samples in the center k-space. 

We note from \eqref{eq:motionupdate} that the motion parameters $\bs \xi$ are estimated only using the measured k-space samples and the forward model that depends on the motion parameters; it does not depend on the missing central k-space samples, and hence the algorithm may not benefit from clever infilling strategies. The alternation between \eqref{eq:cgupdate} and \eqref{eq:denoise} results in the infilling of the missing k-space regions and suppresses noise artifacts.

\begin{figure*}
    \includegraphics[width=\textwidth]{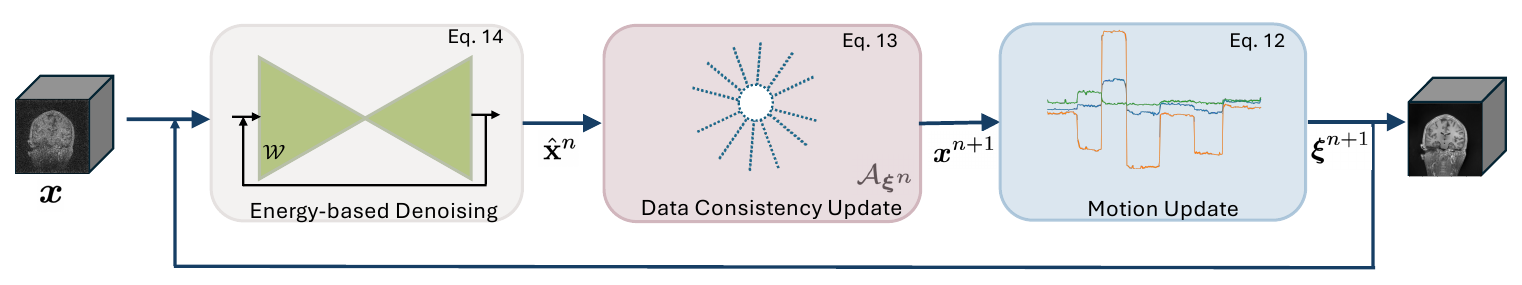}
    \caption{Overview of joint reconstruction and motion correction. We initialize the image volume ${\bs x}$ using CG-SENSE with no motion correction, while the motion parameters $\bs \xi$ are initialized as zeros. The optimization alternates between energy-based denoising specified by Eq.~\eqref{eq:denoise}, data consistency update using CG specified by Eq.~\eqref{eq:cgupdate}, and motion update specified by Eq.~\eqref{eq:motionupdate}. The iterative algorithm recovers the motion-compensated and k-space center-infilled volume. 
    }
    \label{fig:mainfig}
\end{figure*}

\paragraph*{Multi-resolution MoCo}
Prior work suggests that rigid motion can often be estimated reliably from low-resolution images\cite{wu2025moner}. Because the central (low-frequency) k-space region is relatively densely sampled, low-resolution reconstructions derived from these samples tend to be less affected by aliasing and high-frequency inconsistencies than full-resolution images. We use a coarse-to-fine (multi-resolution) motion-estimation strategy. Specifically, we introduce a k-space mask $\mathbf M_s$ that initially retains only a central low-resolution region, which gradually expands to include higher spatial frequencies as the optimization progresses. The motion update in \eqref{motionupdate} is modified to
\begin{eqnarray}
    \bs \xi^{n+1} &=& \bs \xi^n - \alpha\nabla_{\bs \xi}\tfrac{1}{2\sigma^2}\,\big\|\mathbf M_s\big(\bs y-\bs{\mc A}_{\bs \xi}\bs {x}^{n}\big)\big\|_2^2 .
\end{eqnarray}
We start with a k-space mask that corresponds to a resolution of 16\,mm and progressively (16\,mm, 4\,mm, and 1\,mm) enlarge it until it covers the complete sampled region. Importantly, masking is applied only during the motion-parameter update; the image update still uses the full data-consistency step in \eqref{eq:cgupdate}. Fig.~\ref{fig:mainfig}shows the full optimization pipeline.
\begin{algorithm}
    \caption{EnMoCo}
    \label{alg:myalg}
    \begin{algorithmic}[1]

        \State \textbf{Inputs:} measurements $\bs y$, forward operator $\bs{\mc A}$, pretrained $\mc{R}_{\bs{w}}$, other hyper-parameters
        \State \textbf{Outputs:} image $\bs {x}$, motion parameters $\bs \xi$

        \Statex
        \State \textbf{Initialize:} 
        \State $\textit{$\bs \xi$} \gets \text{$\bs 0$}$
        \State $\textit{$\bs {x}$} \gets \text{$\arg \min_{\bs {x}} \|\bs y-\bs{\mc A}_{\bs \xi}\bs {x}\|_2^2 \;+\; \lambda \|\bs {x}\|_2^2$}$

        \Statex
        \For{scale $\in \{16, 4 ,1 \}\ \text{mm}$}
            \State get mask $\bs M_s$ of $scale$
            \For{$n=1$ to $N$}
                \State $\bs \xi^{n} \gets \arg \min_{\bs \xi} \tfrac{1}{2\sigma^2}\,\|\bs M_s(\bs y-\bs{\mc A}_{\bs \xi}\bs {x}^{n-1})\|_2^2$
                \Repeat
                    \State $\hat{\mathbf{x}}^{n} \gets \left(\bs {x}^{(n-1)}-\frac{1}{L}\nabla_{\bs {x}}\mc{R}_{\bs{w}} (\bs {x}^{(n-1)})\right)$
                    \State $\bs {x}^n \gets \left(\frac{\bs{\mc A}_{\bs \xi^{n}}^H \bs{\mc A}_{\bs \xi^{n}}}{\sigma^2}+L\bs I\right)^{-1}\left(\frac{1}{\sigma^2}\bs{\mc A}_{\bs \xi^{n}}^H\bs y+L \hat{\mathbf{x}}^{n}\right)$
                \Until{convergence}
            \EndFor

        \EndFor

        \State \Return $\bs {x}$, $\bs \xi$

    \end{algorithmic}
\end{algorithm}

\subsection{Experiments}
\subsubsection{ZTE Acquisition}
We used a magnetization prepared Zero Echo-time (MP-ZTE) sequence to acquire the $T_1$-weighted images; the sequence is illustrated in Fig.~\ref{suppfig:ir_zte} of the supplementary material. Imaging was performed on a 3T clinical MRI (Signa Premier, GE Healthcare, Waukesha, WI) with a 48-channel receive array head coil (GE Healthcare).
Acquisition parameters for the ZTE-based inversion-recovery scan were 25.6 cm isotropic field of view, matrix size=256, \putmarginpar{R1.C3.}{readout bandwidth=$\pm$31.25 kHz/FOV}, and flip angle=5 degrees. $T_1$-weighted image contrast was achieved using intermittent magnetization preparation pulses at the start of each segmented readout block, each consisting of 192 angularly optimized ZTE free-induction decay radial lines using HEALPix radial view ordering\cite{corum2019healpix}. The delay between the inversion pulse and the first excitation was 1058.9 ms, followed by 192 readouts of TR=2.32 ms, and an 800 ms delay to allow the longitudinal magnetization to recover before the next inversion.  Spokes from all RF excitations were equally weighted in the reconstruction.

Each excitation radiofrequency pulse was followed by the readout block, with the minimum possible delay. \putmarginpar{R1.C3.}{Accounting for hardware switching delays (RF ring-down, coil un-blanking, and signal stabilization) alongside half the duration of the excitation pulse, these constraints combined to a missing k-space radius of $\sim$2.5 readout
samples at a readout bandwidth of $\pm$31.25 kHz/FOV.} Data were acquired from seven normal volunteers who provided informed consent under this IRB-approved study. We acquired one motion-free and one motion-corrupted scan with controlled head motion from each subject. For one of the subjects, two additional motion-corrupted scans were acquired, resulting in four motion levels for that subject. In total, 16 datasets were used for testing. We used two acquisition settings: 406\,s scans with 170 IR segments and 611\,s scans with 256 IR segments. Four datasets from two subjects used the 406\,s acquisition, while the remaining datasets used the 611\,s acquisition. To evaluate the impact of acceleration, retrospective undersampling was performed on the 611\,s acquisitions by retaining partial blocks, e.g., 128 and 64 blocks corresponding to acquisition times of 306 and 153 seconds, respectively. Additionally, we acquired eight motion-free datasets from eight subjects for training.

\subsubsection{Pre-processing}
In this work, we aimed to recover motion parameters $\xi$ of dimension $T\times 6$, where $T$ denotes the number of inversion blocks. We neglected the motion within each inversion block in this study. We used the Stoch--Olejniczak approach \cite{stoch2005missing,corum2020robust} to infill 2.5 samples, from which the coil sensitivity maps were subsequently estimated using JSENSE. The reference image was reconstructed using TV regularization with a weight of 0.002. We simulated gap sizes of 1.5--5 in our experiments, which cover the large range reported in the literature\cite{huoLowrankIterativeInfilling2025}.

\subsubsection{Simulated 3D ZTE Data}
A simulated ZTE dataset derived from MPnRAGE \cite{mpnrage} was used for quantitative validation. We assumed a multichannel non-uniform Fourier transform model mimicking the ZTE acquisition, where Gaussian noise was added to simulate data with four different signal-to-noise ratios (SNR): S1=24.4 dB; S2=19.8 dB; S3=14.0 dB; S4=10.1 dB. Center k-space samples were retrospectively removed to create a range of discrete dead time gaps. Three scan durations were used, where we retained 256, 128, and 64 inversion blocks with acquisition times of 611, 306, and 153 seconds, respectively. The motion parameters used for simulation are shown in Fig.~\ref{fig:simulate_snr_motion}(A).

\begin{figure*}[h]
    \includegraphics[width=\textwidth]{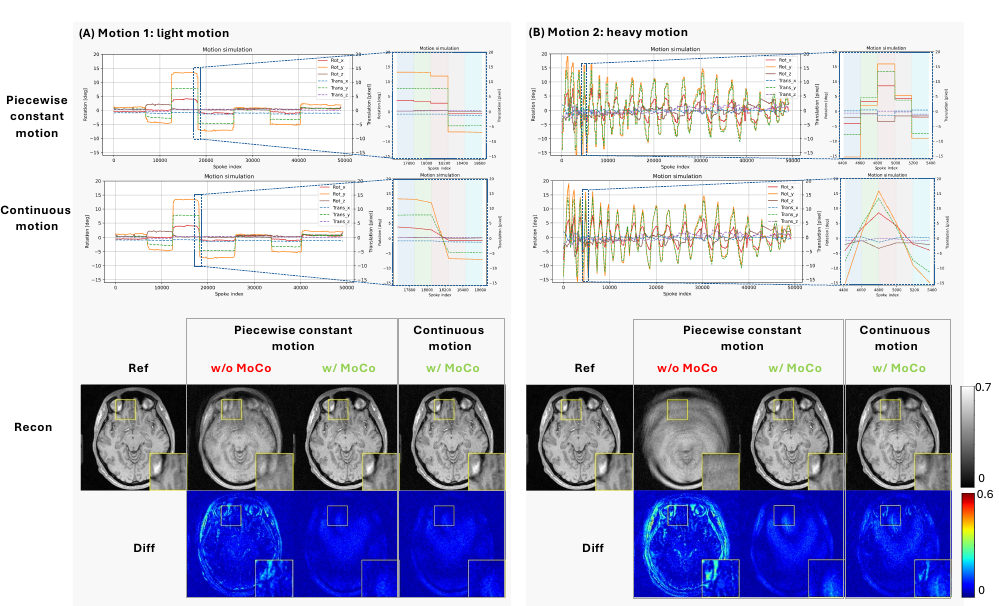}
    \caption{Evaluation of continuous intra-block motion modeling. (A) Motion 1 with light motion and (B) Motion 2 with heavy motion, showing corresponding motion parameters per spoke, reconstructions from EnMoCo with (w/) and without (w/o) applying motion correction (MoCo), and difference maps. Parameter trajectories from five selected blocks are zoomed in and color-coded. Reconstructions from Motion 1 with continuous motion corruption are comparable to piecewise constant motion corruption, demonstrating that the per-block approximation holds well under moderate, realistic continuous head motion. For Motion 2, residual motion artifacts become visible with continuous motion corruption, reflecting localized breakdowns of the block-wise constant assumption during high-velocity movement.}
    \label{fig:intrablockMotion}
\end{figure*}

The proposed method assumed the motion parameters to be the same for all spokes within an inversion block. To assess intra-block motion from continuous movement, an additional simulation was performed. We generated the motion-corrupted data by assuming that the motion parameters were linear across spokes at two different motion levels as shown in Fig.~\ref{fig:intrablockMotion}. The proposed block-wise motion estimation was applied to both profiles to evaluate image quality degradation. 

\subsubsection{Network Architectures and Implementation Details}
 We used a 2D DRUNet model $\mc{D}_{\bs{w}}$ with 64, 128, 256, and 512 channels for four levels, respectively. The model was trained with normalized slices corrupted by noise, with standard deviations $\varsigma$ uniformly chosen from 0 to 0.1. The training dataset included eight motion-free ZTE datasets, acquired and reconstructed using Total Variation (TV), from 611\,s scans. We retrieved 2D slices from the $x$--$y$ and $x$--$z$ planes, discarding those with low SNR, resulting in a total of 2480 valid slices. Each complex 2D slice was treated as a two-channel real image. $\bs{w}$ was trained for 80 epochs using the Adam optimizer with a learning rate of $1\times 10^{-4}$ and a batch size of 32. Training dataset augmentation, including rotation and brightness adjustments, was applied to account for motion and contrast shifts. 
 
 We set the Lipschitz constant $L$ to 2, $\sigma^2$ to 0.1, and the MM exit tolerance to $10^{-3}$. The maximum number of CG iterations was set to 200. An ablation study evaluating hyperparameter tuning is included in Fig.~\ref{suppfig:ablation_lip_cg_outerloop}--\ref{suppfig:ablation_sigma} of the Supporting Information. Three spatial resolutions were adopted for multi-resolution motion estimation: 16\,mm, 4\,mm, and 1\,mm. 50 iterations of motion estimation were adopted in each alternating loop with a learning rate of $0.01$. \putmarginpar{R1.C2.}{All hyperparameters were fixed across all in vivo experiments.} All experiments were conducted on a single NVIDIA A100 GPU with 80\,GB of memory.

\subsubsection{Methods used for Comparisons}
In this section, we describe the baseline methods evaluated for comparison. The proposed approach was compared against ZTE infilling methods both with and without motion correction. CG-SENSE and Low Rank represented state-of-the-art (SOTA) infilling methods without motion correction. For motion-corrected baselines, motion estimation was incorporated into CG-SENSE using either ground-truth motion parameters (CG-SENSE Upper Bound) or motion estimation methods (AF and MERLIN). Additionally, a classical regularization term, Total Variation (TV-MoCo), was integrated alongside motion correction.
 
\paragraph{CG-SENSE} CG-SENSE recovers k-space holes using coil sensitivity maps and quadratic regularization. 
\begin{equation}
\min_{\bs {x}} \|\bs y-\bs{\mc A}\bs {x}\|_2^2 \;+\; \lambda \|\bs {x}\|_2^2.
\end{equation}
Here, we did not account for motion (i.e., $\bs{\xi}=0$) during reconstruction. 

\paragraph{CG-SENSE Upper Bound} 
CG-SENSE Upper Bound (CG-UP) performed a CG-SENSE reconstruction with ground-truth motion parameters on simulated data. It serves as the upper bound for the CG-SENSE when the motion parameters are known. The objective function is 
\begin{equation}
\min_{\bs {x}} \|\bs y-\bs{\mc A}_{\bs \xi}\bs {x}\|_2^2 \;+\; \lambda \|\bs {x}\|_2^2,
\end{equation}
where $\bs \xi$ represents the ground truth motion parameters.

\paragraph{Autofocusing}
Autofocusing (AF) is a motion-correction method that alternates between image updates and motion searches\cite{cordero-grande2020motioncorrected,haskell2018targeted}. We used Tikhonov regularization over the image in this work; the algorithm alternates between the following data consistency updates for the image and motion parameters:
\begin{equation}
    \begin{aligned}
       \bs {x}^{n+1} &= \arg \min_{\bs {x}}\|\bs y-\bs{\mc A}_{\bs \xi^n}\bs {x}\|_2^2 \; + \;\lambda \|\bs {x} \|^2_2\\
       \bs \xi^{n+1} &= \bs \xi^n - \alpha\nabla_{\bs \xi}\|\bs y-\bs{\mc A}_{\bs \xi}\bs {x}^{n+1}\|_2^2.
    \end{aligned}
\end{equation}

\paragraph{MERLIN}
 MERLIN\cite{ljungbergMotionCorrectedSilent2022a} employs image-domain registration to align multiple navigator images to a single template, where a single WASPI acquisition at the start generates coil sensitivity maps. Navigator images from consecutive k-space segments are registered pairwise to the first navigator image, yielding rotation and translation parameters. The corrected k-space data are combined and reconstructed into a high-resolution image, with WASPI data explicitly filling the dead time gap. Inspired by MERLIN's registration-based motion correction framework, we adopted a similar approach: low-resolution navigator images recovered from each segment using SENSE were used to recover motion parameters, followed by motion-compensated reconstruction. Due to the lack of WASPI acquisition in our data, we infilled the central k-space regions implicitly using CG-SENSE. We included this baseline to compare the effectiveness of image-domain registration against the k-space-based motion correction for ZTE neuroimaging.

\paragraph{TV-MoCo}
 We follow an alternating minimization approach similar to Aligned SENSE\cite{cordero2016sensitivity}, where we alternate between Total Variation (TV) regularization and motion-compensated reconstruction; we term this approach TV-MoCo. TV-MoCo alternates between the following steps:
\begin{equation}
    \begin{aligned}
       \bs {x}^{n+1} &= \arg \min_{\bs {x}}\|\bs y-\bs{\mc A}_{\bs \xi^n}\bs {x}\|_2^2 \; + \;\lambda \|\bs G\bs {x} \|_1\\
       \bs \xi^{n+1} &= \bs \xi^n - \alpha\nabla_{\bs \xi}\|\bs y-\bs{\mc A}_{\bs \xi}\bs {x}^{n+1}\|_2^2.
    \end{aligned}
\end{equation}
where $\bs G$ is the gradient operator.

\paragraph{Low Rank Infilling}
The Low Rank iterative infilling method reformulates the finite support constraint in the image domain to the locally low rank property in $k$-space\cite{huoLowrankIterativeInfilling2025}. It also takes advantage of the coil sensitivity map information, similar to the CG-SENSE method. The optimization is posed as follows:
\begin{equation}
    \min_{\bs {x}} \|\bs{\mc A} \bs {x} - \bs y \|^2 + \lambda \|\mathbf{HNS} \bs{x}\|_*.
\end{equation}
where $\bs{\mc A}$ denotes the forward operator excluding the missing samples trajectory, $\mathbf{N}$ denotes the NuFFT operator including the missing samples trajectory, $\mathbf{S}$ refers to the coil sensitivity map, and $\mathbf{H}$ represents the matrix lifting operator to form a structured Hankel matrix from 3D Cartesian data. The high computational complexity of this method makes its extension to motion correction challenging.

\subsubsection{Evaluation Metrics}
\putmarginpar{R1.C3.}{We relied on both qualitative and quantitative metrics for evaluation. In the simulated 3D ZTE settings where ground truth images were available, we evaluated image reconstruction quality using peak signal-to-noise ratio (PSNR) and structural similarity index (SSIM). We used $\ell_1$ distance measures for motion accuracy,
\begin{equation}
    \ell_1(\hat{\bs \xi}) = \frac{1}{T}\sum_{t=1}^T|\hat{\bs \xi} - \bs \xi |,
\end{equation}
where $T$ represents the number of k-space segments. The estimated motion parameters from all data were registered to a common reference image to eliminate linear shift errors.}

\section{Results}
\begin{figure*}
    \centering
    \includegraphics[width=0.5\textwidth]{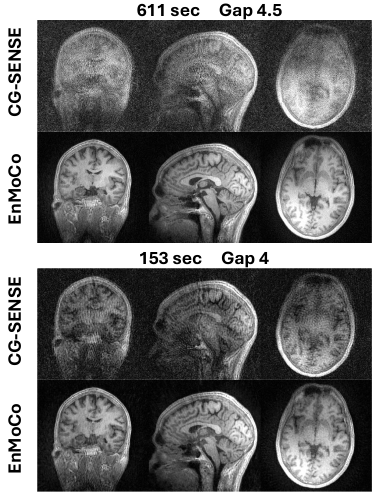}
    \caption{Preview of CG-SENSE and EnMoCo reconstructions from 611\,s and 153\,s motion-corrupted in vivo acquisitions. CG-SENSE reconstruction exhibits blurriness caused by motion in 611\,s data and contrast shift due to acceleration in 153\,s data. EnMoCo resolves motion artifacts using motion correction and reduces contrast shift using MuSE regularization.  
    }
    \label{fig:snapshot}
\end{figure*}

A preview of in vivo reconstructions on 611\,s and 153\,s data is shown in Fig.~\ref{fig:snapshot}. CG-SENSE reconstructions without motion correction suffer from extensive motion artifacts, while EnMoCo is able to significantly reduce these artifacts. In the 153\,s data, CG-SENSE exhibits a contrast shift resulting from missing k-space center and has motion artifacts. EnMoCo is able to reduce the motion artifacts and contrast shift. We perform rigorous experiments on simulated data and in vivo data in the following sections to validate the proposed method.

\subsection{Validation using numerical simulations}
We first evaluate the proposed algorithm on simulated ZTE data to understand the trade-offs and to perform quantitative comparisons with other methods.

\begin{figure*}[h]
    \centering
    \includegraphics[width=0.8\linewidth]{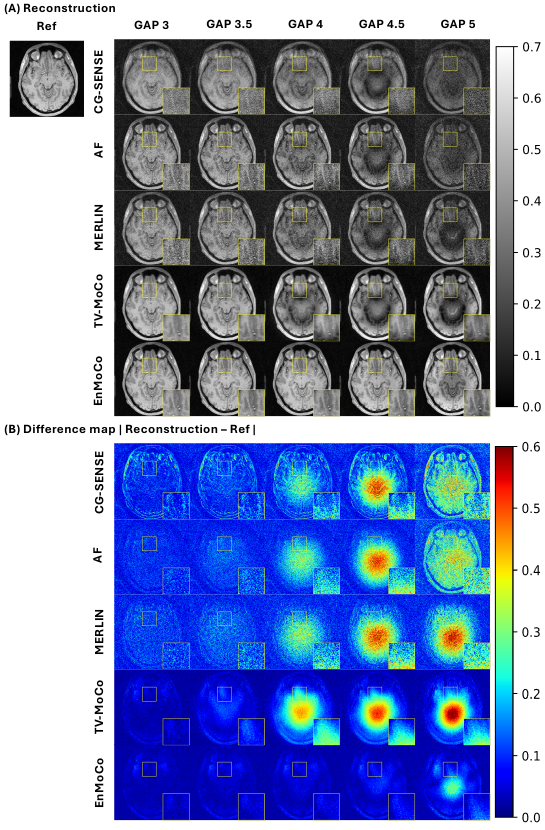}
    \caption{Impact of the dead time gap on motion-compensated recovery in simulated 611\,s ZTE experiments. Reconstructions are shown for dead time gaps of 3–5 samples using CG-SENSE, AF, MERLIN, TV-MoCo, and EnMoCo. CG-SENSE suffers from both aliasing artifacts and motion distortions. MERLIN reduces motion distortions using image-domain registration, while AF further reduces motion distortions. TV-MoCo effectively removes aliasing artifacts; however, contrast shifts caused by missing central samples are amplified at larger gaps (gap $\geq$ 3.5). EnMoCo consistently yields reduced motion and aliasing artifacts, particularly for larger gaps ($\geq$ 4). The PSNR and SSIM are reported in Table~\ref{tab:psnr_1x_simulated}.}
    \label{fig:simulate_1}
\end{figure*}

\begin{figure*}[h]
    \centering
    \includegraphics[width=1\linewidth]{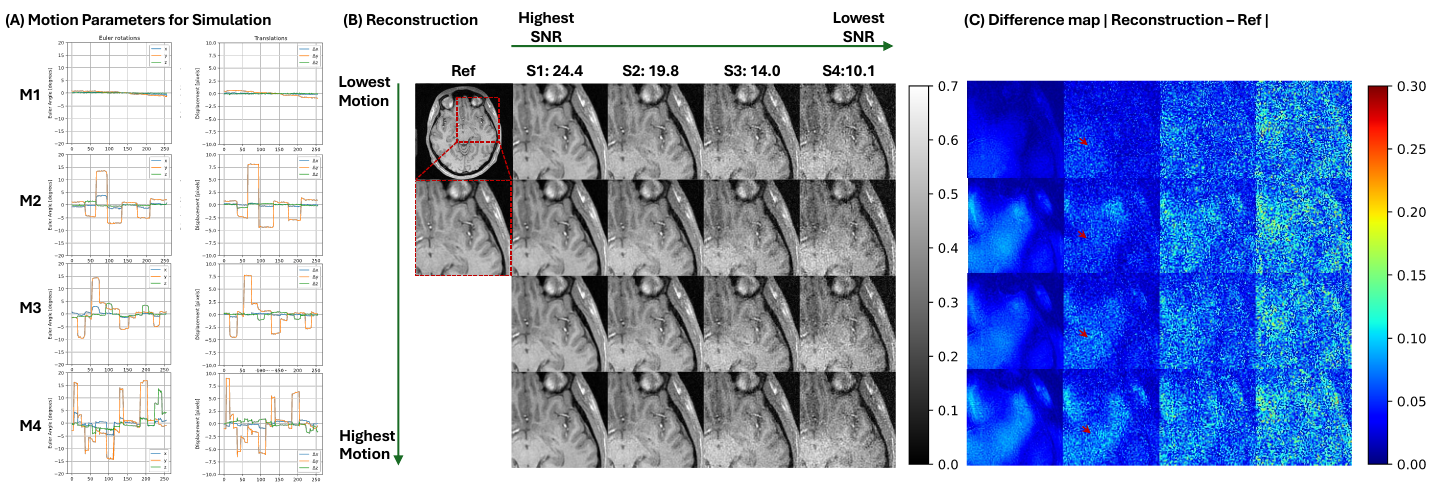}
    \caption{Impact of motion severity and SNR in reconstruction quality on simulated data: Data were simulated at four SNR levels (S1--S4: 24.4--10.1~dB) and four rigid-motion conditions (M1--M4). (A) shows the prescribed block-wise rigid-motion parameters (three rotations and three translations per block) used to generate the corrupted measurements; motion magnitude and/or the number of motion events increase from M1 (mild) to M4 (severe). Reconstructions are shown in (B), and the difference maps in (C). Across SNR levels, the reconstructions are observed to remain visually sharp under mild-to-moderate motion, indicating that the motion updates are relatively robust to SNR. At a fixed SNR, the difference maps show a moderate increase in residual errors and noise variance with increasing motion severity. We note that motion alters the effective location of the k-space spokes, resulting in some regions of the k-space sphere being undersampled; this translates to error amplification and aliasing artifacts, as indicated by red arrows in (C). We observe that MuSE regularization keeps the increase in noise variance moderate. By contrast, Fig. \ref{fig:simulate_1}shows that AF exhibits significantly higher noise amplification even at M2. }
    \label{fig:simulate_snr_motion}
\end{figure*}

\paragraph{Impact of dead time gap}  Fig.~\ref{fig:simulate_1} illustrates the impact of dead time gap under intermediate motion (M2: $\pm 15^\circ$, $\pm 8$~mm) and an SNR of 19.8~dB (S2 from Fig.~\ref{fig:simulate_snr_motion}). CG-SENSE partially compensates for missing data, restoring image contrast for dead time gaps, but it exhibits motion-induced blurring. MERLIN (CG-SENSE + registration-based motion estimation) reduces motion distortions using image-domain registration, while AF (CG-SENSE + k-space motion estimation) further improves motion estimation accuracy. TV-MoCo effectively removes aliasing artifacts; however, contrast shifts caused by missing central samples are amplified at larger gaps (gap $\geq$ 3.5). EnMoCo consistently yields reduced motion and aliasing artifacts, particularly for larger gaps ($\geq$ 4). EnMoCo effectively suppresses motion and aliasing artifacts by combining motion estimation with an energy-based denoiser. The reduction in errors is also evident in the difference map. We attribute the gains to the improved regularization provided by MuSE. The results of the gaps below 3 and other methods, including Low Rank and CG-UP, are included in Fig.~\ref{suppfig:simulate_1x_allmethods} and Table~\ref{supptab:psnr_1x_simulated} of the Supporting Information. Table~\ref{supptab:l1_1x_simulated} summarizes the $\ell_1$ motion estimation errors for AF and EnMoCo. Fig.~\ref{suppfig:ablation_study} presents an ablation study evaluating the relative contributions of energy-based regularization and motion correction.

\paragraph{Runtime analysis}
The execution times across varying gap sizes are summarized in Table~\ref{tab:runtime}. Conventional methods (CG-SENSE, AF, and MERLIN) rely on standard conjugate-gradient iterations, resulting in execution times that remain largely invariant to gap size. By contrast, TV-MoCo and Low Rank involve sharp increases in runtime as the gap increases. For example, the complexity of the Low Rank method \cite{huoLowrankIterativeInfilling2025} increases rapidly from 3~min at a gap of 1.5 to 920~min at gaps $\ge 4$, while TV-MoCo reaches up to 247~min at a gap of 5 due to degraded convergence under severe central k-space loss. We note that the runtime of EnMoCo is higher for small gaps (28~min); its runtime scales far more gracefully as the gap size expands (reaching only 40~min at a gap of 5).

\paragraph{Impact of motion severity and SNR}
 Fig.~\ref{fig:simulate_snr_motion}summarizes results for a constant gap size of 4 samples, four SNR levels (S1--S4: 24.4--10.1~dB), and four rigid-motion conditions with increasing motion severity (M1--M4: $\pm 2^\circ$ to $\pm 17^\circ$). 
 
The reconstructed images remain sharp across motion levels at all SNR levels, indicating that motion correction is less sensitive to noise (Fig.~\ref{fig:simulate_snr_motion}(B)). At a fixed SNR, increasing motion severity results in relatively slow rises in residual errors in the difference maps shown in Fig.~\ref{fig:simulate_snr_motion}(C). A sharp increase in errors would be expected due to effective non-uniform k-space sampling from spoke rotations caused by motion, leading to error amplification and alias artifacts. This degradation is minimized by the MuSE regularizer, which controls noise amplification and artifacts. From Fig.~\ref{fig:simulate_1}, we note that at a gap size of 4, an SNR of S2:~19.8~dB, and motion level M2, AF exhibits significantly more noise amplification, indicating greater sensitivity of SENSE to noise and/or imperfect conditioning. Table~\ref{supptab:l1_snr_motion} and \ref{supptab:psnr_snr_motion} report the $\ell_1$ error of motion estimation and PSNR/SSIM across multiple SNR and motion levels. \putmarginpar{R2.C1.}{An ablation study investigating the robustness of the 2.5D prior to rotation is presented in Fig.~\ref{suppfig:ablation_rotation} of the Supporting Information. Rotation about the $x$--axis exhibits a slight degradation compared to $y$-- and $z$--axis rotations at larger angles, as $x$--axis rotation causes out-of-plane motion across both the $x$--$y$ and $x$--$z$ training planes, whereas $y$-- and $z$--axis rotations maintain in-plane transformations for one of the trained planes.}

\begin{figure*}[h]
    \centering
    \includegraphics[width=1\linewidth]{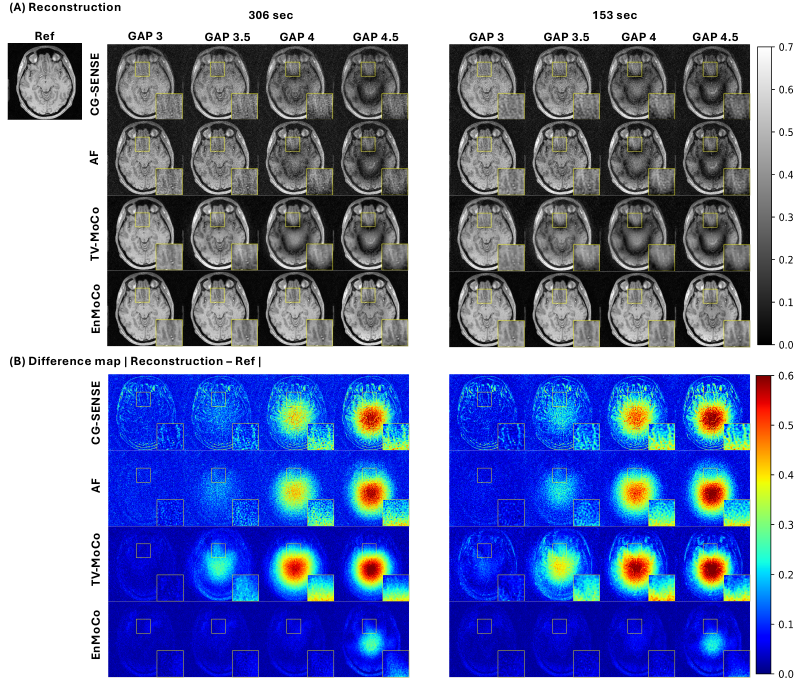}
    \caption{Impact of scan acceleration and dead time gap size on simulated image quality and motion estimation. Reconstructions are shown for undersampled acquisitions retaining 128 blocks ($306\,\text{s}$) and 64 blocks ($153\,\text{s}$) out of 256 full inversion blocks across gap sizes ranging from $3.0$ to $4.5$. CG-SENSE and AF exhibit severe aliasing artifacts and contrast distortion due to missing central k-space data, particularly for gaps $> 3.5$. TV-MoCo exhibits motion distortion at the $153\,\text{s}$ scan duration. In contrast, EnMoCo consistently suppresses motion artifacts and recovers missing k-space samples up to gap size $4.0$, demonstrating superior robustness under high acceleration.}
    \label{fig:simulate_2x_4x}
\end{figure*}

\paragraph{Impact of acceleration}
Fig.~\ref{fig:simulate_2x_4x} evaluates performance under accelerated 3D ZTE acquisitions by retrospectively reducing the number of inversion blocks to 128 and 64, corresponding to scan durations of 306\,s and 153\,s, respectively. 

At the 153\,s scan duration, AF exhibits pronounced contrast deviations and aliasing even at a gap size of 3.5, which worsen at a gap of 4. While TV-MoCo provides reasonable artifact suppression at 306\,s, its performance degrades substantially on the 153\,s dataset: the reduced data redundancy destabilizes its motion parameter estimation, leaving visible residual blurring and localized structural errors. In contrast, EnMoCo maintains high-fidelity reconstructions across gap sizes up to 4 across both acceleration factors. Only mild contrast degradation is observed at the extreme gap size of 4.5 under the 153\,s setting, highlighting the stabilizing role of the learned prior in joint motion and image estimation under severe data undersampling. PSNR and SSIM are summarized in Table~\ref{supptab:psnr_2x_4x_simulated} of the Supporting Information. Table~\ref{supptab:l1_2x_4x_simulated} summarizes the $\ell_1$ motion estimation errors.

\paragraph{Impact of intra-block motion}
Under moderate, realistic continuous head motion (Motion 1 in Fig.~\ref{fig:intrablockMotion}), residual blurring and artifact levels are negligible relative to reconstructions without intra-block corruption, demonstrating that the block-wise constant approximation holds remarkably well in typical practical scenarios. However, under severe continuous motion (Motion 2), residual motion artifacts become visible in difference maps, reflecting localized breakdowns of the piecewise constant assumption during high-velocity movement. Fig.~\ref{suppfig:matchcondition} in the Supporting Information shows a comparison between simulation and in vivo data with more motion patterns. The results confirm that performance degradation caused by continuous intra-block motion remains minimal for typical slow-to-moderate motion.  

\subsection{In vivo Analysis}

\begin{figure*}[h]
    \centering
    \includegraphics[width=0.85\linewidth]{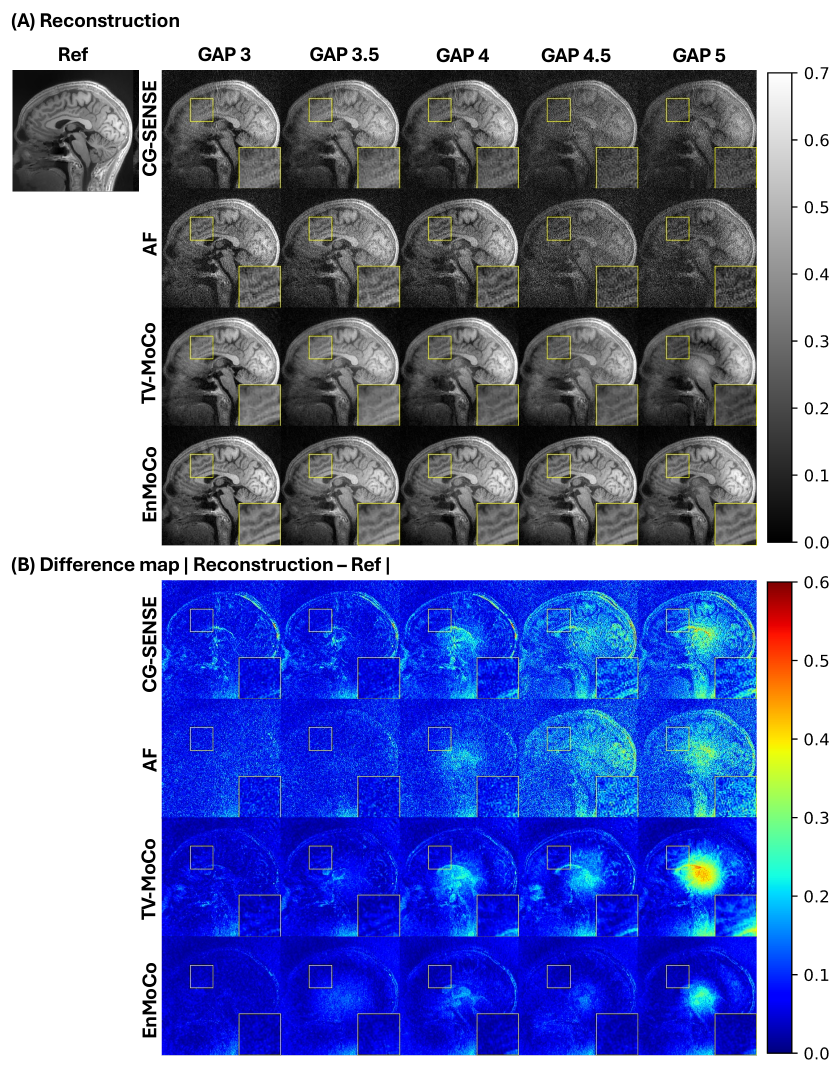}
    \caption{Evaluation of missing samples infilling and motion correction on in vivo 611\,s ZTE data. The motion level is comparable to M2 shown in Fig. \ref{fig:simulate_snr_motion}. We observe that CG-SENSE preserves global image contrast, but exhibits pronounced structural artifacts resulting from uncorrected motion. AF mitigates motion artifacts for gaps up to 4, but suffers from blurriness and contrast inconsistencies as the number of missing samples increases. TV-MoCo performs well on gaps of 3 and 3.5, but exhibits contrast shift for higher gap sizes. The proposed method further improves image quality, reducing both structural artifacts and contrast shift. However, the image quality is less effective than in the simulated experiments, which we attribute to the imperfections in the reference image, inaccuracies in the coil sensitivity maps, intra-block motion, and the inherently lower SNR of real data. }
    \label{fig:real_1x}
\end{figure*}

\paragraph{Comparison with SOTA methods}
 Fig.~\ref{fig:real_1x} compares reconstructions on in vivo ZTE data across gap sizes 3 to 5 ($\pm 15^\circ$, $\pm 10$~mm). CG-SENSE exhibits severe noise and motion blur as the gap size increases. AF mitigates motion artifacts up to gap 4, but suffers from blurring and contrast loss at larger gaps. TV-MoCo improves artifact suppression at small gaps, but degraded motion estimates at gaps $\ge 4.5$ introduce residual blur. In contrast, EnMoCo maintains stable image quality across nearly all gap sizes, effectively suppressing artifacts and preserving contrast. Comparisons of accelerated in vivo data and multiple motion levels are included in Fig.~\ref{suppfig:real_multimotion}--\ref{suppfig:real_2x_4x} of the Supporting Information. Across various acceleration factors, motion severity levels, and dead time gaps, the proposed method robustly suppresses the majority of motion artifacts while preserving fine anatomical structures. Fig.~\ref{suppfig:ablation_mocoscale} in the Supporting Information demonstrates the advantage of multi-scale over single-scale motion correction, showing improved sharpness on in vivo data.

\begin{figure*}[!ht]
    \centering
    \includegraphics[width=0.95\linewidth]{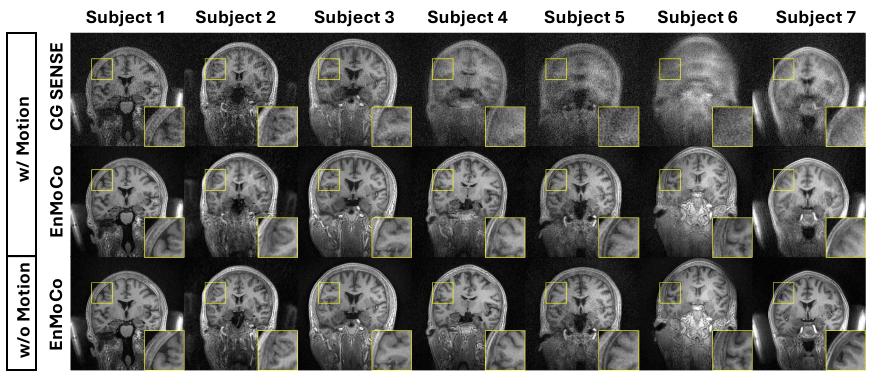}
    \caption{Evaluation on seven subjects. EnMoCo improves the sharpness of the image reconstructed from corrupted data. EnMoCo successfully reconstructs high-quality ZTE images across a wide range of motion levels, from minimal, e.g., Subject 1, to extreme motion, e.g., Subject 7. By contrast, CG-SENSE is corrupted by motion artifacts for scans with heavy motion, e.g., Subjects 4 -- 7. For motion-free acquisitions, EnMoCo still performs motion estimation but does not degrade the reconstruction quality. 
    }
    \label{fig:newsubjects}

\end{figure*}

\paragraph{Qualitative evaluation on multiple subjects}
 Fig.~\ref{fig:newsubjects}compares: CG-SENSE reconstruction of the motion-corrupted scan (no motion correction), EnMoCo of the motion-corrupted scan, and EnMoCo of the motion-free scan from seven subjects with a gap size of 3.

The extent of motion varies across subjects, as reflected in the uncorrected CG-SENSE reconstructions. Across subjects, EnMoCo reduces visible motion-related artifacts in the motion-corrupted scans and does not introduce obvious degradation in the motion-free scans. In cases with more pronounced motion (e.g., Subjects 6 and 7), residual blurring remains after correction, although the reconstructed images show improved sharpness and reduced artifact burden relative to the uncorrected reconstructions. Fig.~\ref{suppfig:newsubjects} in the Supporting Information shows additional axial and sagittal reconstruction views.

\section{Discussion}
This work presents a unified reconstruction framework for 3D ZTE that jointly addresses dead-time–induced missing central k-space samples, rigid subject motion, and undersampling  within a regularized MAP formulation. Regularization is provided by a pre-trained Multi-Scale Energy-based model (MuSE), used as an explicit energy and optimized using a majorization-minimization strategy, which is well-suited to the ill-conditioning introduced by missing-center data and acceleration. To reduce 3D training demands, we use a data-efficient 2.5D construction that aggregates slice-wise energies from two orthogonal views, aiming to mitigate view-dependent artifacts while remaining practical in limited-data settings. 

Several methodological and practical considerations merit discussion for future work. First, hyperparameter selection required careful balancing across varying gap sizes and motion levels; while fixed parameters performed robustly under moderate conditions, adaptive tuning could further optimize extreme cases. Second, evaluating in vivo reconstructions is inherently constrained by the absence of a gold-standard reference. Moreover, hyperparameters tuned on synthetic data may not be optimal for in vivo datasets. This performance gap between simulated and experimental data primarily stems from lower signal-to-noise ratio, imperfect coil sensitivity maps, and continuous off-resonance effects.

Our implementation assumes rigid-body motion and neglects intra-block motion during one inversion block. This assumption may break down under severe, continuous, high-velocity movement. Consequently, improving the temporal resolution of motion correction remains an important direction for future investigation. Directly comparing our approach with the full MERLIN\cite{ljungbergMotionCorrectedSilent2022a} pipeline using WASPI infilling is left for future work, as this requires a dedicated acquisition protocol not retrospectively available in our current dataset. \putmarginpar{R2.C1.}{The 2.5D MuSE prior substantially reduces training data and memory requirements compared with a full 3D prior, but it is trained only with in-plane rotation augmentation. Augmenting the training data with slices extracted from 3D volumes rotated out-of-plane could improve the robustness of the 2.5D prior under extreme motion.}

Finally, this work does not account for individual spoke-dependent $T_1$-weighting variations during the acquisition of multiple FIDs that result in small effective TI changes over the course of each inversion recovery. Because the $T_1$ recovery curve is smooth and central k-space dominates overall image contrast, the effective $T_1$ weighting represents a blended average across all 192 readouts rather than a single discrete TI. \putmarginpar{R1.C3.}{A more principled approach is to explicitly resolve the TI variation along the readout train, reconstructing images at multiple inversion times as in MPnRAGE\cite{mpnrage} and performing motion correction on these TI-resolved images, as demonstrated for cardiac $T_1$ mapping\cite{becker2020fast}. Such $T_1$-resolved motion correction would avoid contrast mismatches between the static template image and spokes acquired across a wide inversion time range.}

\putmarginpar{R1.C3.}{Although evaluated on ZTE, the framework is not ZTE-specific. The k-space motion model and the MuSE prior apply directly to conventional 3D radial trajectories, which correspond to the special case of no dead time gap. Because the k-space center is measured directly, we expect the image update to be better conditioned and to converge faster. Validation on non-ZTE 3D radial data remains future work.}

\section{Conclusion}
We introduced a unified MAP-based reconstruction method for accelerated 3D ZTE that jointly estimates the image and rigid-motion parameters while implicitly accounting for unmeasured k-space samples arising from dead time and undersampling. The approach alternates k-space motion updates with conjugate-gradient data-consistency steps and MuSE-based regularization using an explicit energy prior in a data-efficient 2.5D form.
Results on simulated and experimental datasets demonstrate improved robustness to motion and missing-center effects relative to the tested baselines, with remaining failure modes primarily at the largest dead time gaps and highest motion/acceleration settings. These findings suggest that combining explicit k-space motion modeling with an energy-based learned prior is a practical direction for improving motion robustness and image quality in accelerated ZTE acquisitions.

\subsection*{Acknowledgments}
The work is supported by R01-AG067078 and R01-EB019961. The data acquisition was supported by R43MH122028 and was acquired on an MRI instrument funded by S10OD025025. The University of Iowa receives research support from GE Healthcare. University of Virginia receives support from Siemens Healthineers and Canon Medical Systems.

\bibliography{MRM-AMA}%
\vfill\pagebreak

\section*{Supporting information}

The following supporting information is available as part of the online article:

\vskip\baselineskip\noindent

\noindent

\supptabcaption{psnr_1x_simulated}{PSNR (dB)/SSIM of simulated 611\,s ZTE data with varied gaps. All reconstructions are registered to the reference image. TV-MoCo and EnMoCo outperform other methods across gap sizes with spatial regularization and k-space motion correction. EnMoCo recovers images with the highest PSNR and SSIM for gaps $>$ 3, demonstrating the superior performance of a learned prior at larger gap sizes.}

\supptabcaption{psnr_2x_4x_simulated}{PSNR (dB)/SSIM of simulated 306\,s and 153\,s ZTE data with gaps 3 -- 4.5. EnMoCo maintains high-fidelity reconstructions across gap sizes up to 4 across both acceleration factors. Note that the motion of the 306\,s acquisition is heavier than the 153\,s data due to the variation of motion across blocks, as reflected by the PSNR difference between CG-UP for both acquisition times.}
\supptabcaption{l1_1x_simulated}{$\ell_1$ Motion error (mean $\pm$ std) of simulated 611\,s ZTE data. Rotation (degrees) and Translation (pixels). EnMoCo achieves lower $\ell_1$ error for rotation parameters, which we attribute to multi-resolution search. The $\ell_1$ error of translation from both AF and EnMoCo is higher than that of rotation, as shown in Fig.~\ref{suppfig:motion_snr_motionlevel}. Small rotational perturbations are easier to detect and estimate accurately, as they induce pronounced motion artifacts in 3D MRI. }

\supptabcaption{l1_2x_4x_simulated}{$\ell_1$ motion error (mean $\pm$ std) of simulated 306\,s and 153\,s ZTE data. Rotation in degrees and translation in pixels. EnMoCo achieves substantially lower rotation errors than AF for both accelerated acquisitions, while translation errors are comparable.}

\supptabcaption{l1_snr_motion}{$\ell_1$ error of motion estimation with varying SNR and motion levels: Four motion levels, M1--M4, are simulated with increasing motion amplitude and number of movements across inversion blocks, as shown in Fig.~\ref{suppfig:motion_snr_motionlevel}. Upper: Rotation (degrees), Lower: Translation (pixels). The $\ell_1$ error degrades gracefully as the SNR decreases.}

\supptabcaption{psnr_snr_motion}{PSNR (dB)/SSIM for simulated ZTE data with varying SNR and motion levels. Four SNR levels are simulated, ranging from 24.4 dB to 10.1 dB. Four motion levels, M1--M4, are simulated with increasing motion amplitude and number of movements across inversion blocks. Simulated motion parameters are shown in Fig.~\ref{suppfig:motion_snr_motionlevel}. At a given SNR level, reconstructed image quality is primarily governed by the severity of motion. Notably, a single set of hyperparameters was utilized across all SNR and motion levels; consequently, the reported metrics reflect fixed parameter performance rather than individually optimized global solutions.}

\noindent
\suppfigcaption{motion_snr_motionlevel}{Motion estimation from simulated ZTE with varying SNR and motion corruption. Small rotational perturbations lead to pronounced motion artifacts in 3D MRI, making them highly detectable and easier to estimate accurately. In contrast, translations often produce subtler effects, resulting in less precise estimation.}

\suppfigcaption{ir_zte}{Details of ZTE sequence: (a) Magnetization prepared Zero Echo-time (MP-ZTE) sequence. We use a non-selective inversion pulse, followed by a delay to obtain the desired $T_1$ weighting. (b) We use 192 $5^\circ$ excitations within each individual inversion segment. Gradients increment between RF pulses. (c) HEALPix segment gradient values and view endpoints for 4$\times$192 view segments (768 views). (d) Diagram of the outer point from each k-space segment demonstrating relatively uniform angular coverage of k-space.}

\suppfigcaption{simulate_1x_allmethods}{Comparisons from 611\,s simulated data across gaps 1.5-5. Low Rank effectively reduces the artifacts and contrast shift but exhibits blurring artifacts caused by motion. While Low Rank and TV-MoCo suppress noise more effectively than CG-SENSE, their computational runtimes scale sharply with increasing gap size. By contrast, EnMoCo consistently yields reduced motion and aliasing artifacts while maintaining favorable runtime scaling as the gap size expands (Table~\ref{tab:runtime} in the manuscript).}

\suppfigcaption{matchcondition}{Comparison of simulated and in vivo experiments under realistic motion and noise conditions. (A) Impact of motion pattern and intra-block motion: Reconstruction results across increasing motion severities (light to heavy) for (a) in vivo acquisitions, (b) synthetic data simulated with piecewise-constant per-block motion, and (c) synthetic data simulated with continuous intra-block motion. (d) Corresponding rotation and translation trajectories extracted from in vivo datasets. Continuous motion simulation (c) reproduces the motion-dependent blurring observed in vivo, whereas piecewise-constant simulation (b) underestimates degradation under heavy, rapid motion. (B) Impact of signal-to-noise ratio (SNR): Synthetic reconstructions evaluated across decreasing SNR levels under piecewise-constant and continuous motion models. Decreasing SNR combined with continuous motion modeling closes the performance gap between synthetic simulations and in vivo observations.}

\suppfigcaption{ablation_study}{Ablation study evaluating regularization strategy and motion correction components. Reconstruction results, PSNR/SSIM metrics, and absolute difference maps relative to the motion-free reference (Ref):(A)\,vs.\,(B) Regularization Impact (MuSE vs. Tikhonov): The proposed MuSE prior significantly reduces noise and gap-infilling artifacts compared to standard Tikhonov-regularized CG-SENSE. (A)\,vs.\,(C) Regularization Impact (MuSE vs. Total Variation): The MuSE prior yields superior artifact suppression and detail preservation over TV regularization. (A)\,vs.\,(D) Joint Reconstruction vs. Image-Only: Reconstructing without the motion correction step ( EnMoCo w/o MoCo) results in a $7.1\text{ dB}$ PSNR drop ($25.5\text{ dB}$), confirming that image estimation and motion correction are tightly coupled and must be solved jointly.}

\suppfigcaption{ablation_mocoscale}{Ablation study comparing multi-scale and single-scale MoCo on the 611\,s in vivo data. Multi-scale MoCo incorporates spatial resolutions of 16~mm, 4~mm, and 1~mm. Single-scale MoCo applies optimization at 1~mm. Compared to single-scale MoCo, which exhibits residual blurring, multi-scale MoCo achieves sharper reconstruction of subtle brain structures and superior preservation of fine details.}

\suppfigcaption{ablation_lip_cg_outerloop}{Ablation study of Lipschitz constants and outer loop steps. A smaller Lipschitz constant provides faster convergence but suffers from instability, e.g., Lipschitz=0.8. We adopt a Lipschitz constant equal to 2 to balance convergence stability and speed. We evaluate the global loss and the accuracy of motion estimation across 10 outer iterations. Both the global loss and motion estimation converge within 10 outer iterations.}

\suppfigcaption{ablation_sigma}{Ablation study of $\sigma^2$. $1/2\sigma^2$ determines the weights of data consistency and prior. Lower $\sigma^2$ indicates higher importance of data consistency. We observe that the reconstructed image from $\sigma^2=0.01$ is corrupted by aliasing artifacts due to an insufficient prior term. $\sigma^2=1$ is over-regularized by higher weights of the prior term and is not consistent with measurements, as shown by the lower PSNR. We choose $\sigma^2=0.1$, which achieves a better trade-off between data consistency and prior.}

\suppfigcaption{ablation_rotation}{Robustness evaluation of the 2.5D prior across rotation angles ($0^\circ$–$30^\circ$) along the $x$-, $y$-, and $z$-axes. EnMoCo demonstrates stable reconstruction performance (measured in PSNR) for rotation angles between $0^\circ$ and $20^\circ$, covering typical clinical motion regimes and the in vivo data tested in this study. Rotation about the $x$-axis exhibits a slight degradation compared to $y$- and $z$-axis rotations at larger angles, as $x$-axis rotation causes out-of-plane motion across both the $x$-$y$ and $x$-$z$ training planes, whereas $y$- and $z$-axis rotations maintain in-plane transformations for one of the trained planes.}

\suppfigcaption{real_multimotion}{Evaluation of reconstructions under four levels of motion corruption. CG-SENSE reconstruction without motion correction exhibits increasing blurriness and structural degradation as motion severity increases. The proposed method suppresses the majority of motion artifacts across different motion levels and dead time gaps. Notably, we observe that the image quality is more sensitive to the dead time gap than to the motion level. For a gap of 5, severe motion leads to noticeable contrast shifts and structural loss. When the gap is 3, the motion-induced degradation is substantially reduced. }

\suppfigcaption{real_2x_4x}{Results of accelerated in vivo ZTE acquisition. We retain 128 and 64 inversion blocks out of 256, corresponding to a scan time of 306 and 153 seconds, respectively. The dead time gap ranges from 3 to 4.5. Aliasing artifacts dominate the image quality at higher acceleration, as visible in the AF result shown above. The performance of the proposed method is robust, with improved anatomical structures for different acceleration factors.}

\suppfigcaption{newsubjects}{Evaluation on seven subjects showing coronal (A), sagittal (B), and axial (C) reformats of the 3D volumes. Two acquisitions were obtained from each subject, with and without intentional motion. EnMoCo improves the sharpness of the image reconstructed from corrupted data. EnMoCo successfully reconstructs high-quality ZTE images across a wide range of motion levels, from minimal, e.g., Subject 1, to extreme motion, e.g., Subject 7. By contrast, CG-SENSE is corrupted by motion artifacts for scans with heavy motion, e.g., Subjects 4 -- 7. For motion-free acquisitions, EnMoCo still performs motion estimation but does not degrade the reconstruction quality.  }

\vspace*{6pt}


\begin{table*}[!ht]
  \centering
  \caption{PSNR (dB)/SSIM of simulated 611\,s ZTE data with varied gaps. All reconstructions are registered to the reference image. Results of Low Rank and CG-UP are included in the Supporting Information Table~S1. }
  \setlength{\tabcolsep}{4pt}
    \begin{tabular}{c|ccccc}
    \hline
    Methods\textbackslash{}Gaps & 3   & 3.5     & 4   & 4.5    & 5 \\
    \hline
   
    Gridding & 16.9/0.460 & 15.9/0.434 & 15.6/0.422 & 15.4/0.411 & 14.8/0.401 \\
    \hline
    
    CG-SENSE  & 20.9/0.292 & 20.5/0.281 & 19.0/0.259 & 17.2/0.244 & 15.3/0.145 \\
    \hline
    
    AF    & 22.6/0.415 & 21.9/0.402 & 19.7/0.372 & 17.8/0.346 & 15.4/0.200 \\
    \hline
    
    MERLIN & 21.5/0.345 &	20.8/0.332 & 18.9/0.305 & 17.2/0.287 &16.6/0.264 \\
    \hline

    TV-MoCo & 32.8/\textbf{0.841} & 30.4/0.826 & 21.4/0.743 & 20.6/0.728 & 19.8/0.693 \\
    \hline
    
    EnMoCo & \textbf{33.6}/0.839 & \textbf{32.8/0.841} & \textbf{32.6/0.829} & \textbf{32.4/0.824} & \textbf{28.4/0.805} \\
    \hline
    
    \end{tabular}%
  \label{tab:psnr_1x_simulated}%
\end{table*}%

\begin{table*}[h!]
\centering
\caption{Runtime (in minutes) for each method across different gap sizes from a 611\,s acquisition. Conventional methods (CG-SENSE, AF, and MERLIN) rely on standard conjugate-gradient iterations, resulting in execution times that remain largely invariant to gap size. TV-MoCo and Low Rank involve sharp increases in runtime as the gap increases. The runtime of EnMoCo scales far more gracefully as the gap size expands.}
\label{tab:runtime}
\begin{tabular}{l *{7}{c}}
\toprule
\multirow{2}{*}{\textbf{Method}} & \multicolumn{7}{c}{\textbf{Gap Size}} \\
\cmidrule(lr){2-8}
& \textbf{1.5} & \textbf{2.5} & \textbf{3} & \textbf{3.5} & \textbf{4} & \textbf{4.5} & \textbf{5} \\
\midrule
    CG-SENSE        & \textbf{1.4}  & \textbf{1.5}  & \textbf{1.6}   & \textbf{1.8}  & \textbf{2.0}   & \textbf{2.5} & \textbf{2.6} \\ 
    Low Rank         & 3 & 33 & 101 & 340  & 920 & 920 & 920 \\
    AF              & 10  & 11 & 11  & 11  & 11 & 12  & 12 \\
    MERLIN          & 11   & 11   & 11    & 11   & 11  & 12  & 12 \\
    
    TV-MoCo         & 6    & 8    & 38    & 181  & 183 & 239 & 247 \\
    
    EnMoCo       & 28   & 32   & 32    & 35   & 37  & 40  & 40 \\
    \bottomrule
    \end{tabular}
\end{table*}
\end{document}